\documentclass[aps,prb,twocolumn,superscriptaddress,]{revtex4-1}
\usepackage{graphicx}
\usepackage{dcolumn}
\usepackage{bm}
\usepackage{amsmath}
\usepackage{hyperref}
\usepackage[english]{babel}
\usepackage[utf8]{inputenc}
\usepackage{xr}
\usepackage{float}
\def\urlprefix{Preprint at }
\begin{document}
\setlength{\parskip}{0pt}
\bibliographystyle{naturemag_edit}
\title{Low disordered, stable, and shallow germanium quantum wells: a playground for spin and hybrid quantum technology}

\author{A. Sammak}
\thanks{These two authors contributed equally}
\affiliation{QuTech and Kavli Institute of Nanoscience, TU Delft, P.O. Box 5046, 2600 GA Delft, The Netherlands}
\affiliation {QuTech and TNO, Stieltjesweg 1, 2628 CK Delft, The Netherlands}
\author{D. Sabbagh}
\thanks{These two authors contributed equally}
\author{N.W. Hendrickx}
\author{M. Lodari}
\author{B. Paquelet Wuetz}
\author{L. Yeoh}
\affiliation{QuTech and Kavli Institute of Nanoscience, TU Delft, P.O. Box 5046, 2600 GA Delft, The Netherlands}
\author{M. Bollani}
\affiliation{IFN–CNR, LNESS, Via Anzani 42, 22100 Como, Italy}
\author{M. Virgilio}
\affiliation{Dipartimento di Fisica ``E. Fermi'', Universit\`a di Pisa,  Largo Pontecorvo 3, 56127 Pisa, Italy}
\author{M. A. Schubert}
\author{P. Zaumseil}
\affiliation{IHP, Im Technologiepark 25, 15236 Frankfurt}
\author{G. Capellini}
\affiliation{IHP, Im Technologiepark 25, 15236 Frankfurt}
\affiliation{Dipartimento di Scienze, Universit\`a degli studi Roma Tre, Viale Marconi 446, 00146 Roma, Italy}
\author{M. Veldhorst}
\author{G. Scappucci}
\email{g.scappucci@tudelft.nl}
\affiliation{QuTech and Kavli Institute of Nanoscience, TU Delft, P.O. Box 5046, 2600 GA Delft, The Netherlands}
\date{\today}
\def\bibsection{\section*{\refname}} 
\begin{abstract}
Buried-channel semiconductor heterostructures are an archetype material platform to fabricate gated semiconductor quantum devices. Sharp confinement potential is obtained by positioning the channel near the surface, however nearby surface states degrade the electrical properties of the starting material. In this paper we demonstrate a two-dimensional hole gas of high mobility ($5\times 10^{5}$ cm$^2$/Vs) in a very shallow strained germanium channel, which is located only 22 nm below the surface. This high mobility leads to mean free paths $\approx6 \mu m$, setting new benchmarks for holes in shallow FET devices. Carriers are confined in an undoped Ge/SiGe heterostructure with reduced background contamination, sharp interfaces, and high uniformity. The top-gate of a dopant-less field effect transistor controls the carrier density in the channel. The high mobility, along with a percolation density of $1.2\times 10^{11}\text{ cm}^{-2}$, light effective mass (0.09 m$_e$), and high g-factor (up to $7$) highlight the potential of undoped Ge/SiGe as a low-disorder material platform for hybrid quantum technologies.
\end{abstract}
\maketitle
\setlength{\textwidth}{183mm}
\section*{Introduction}
Germanium (Ge) has the highest hole mobility of common semiconductors and is integrated onto Si substrates within a foundry-qualified process\cite{Pillarisetty2011academic}. These properties make high-speed Ge transistors appealing for extending chip performance in classical computers beyond the limits imposed by miniaturization. Ge is also emerging as a promising material for quantum technology as it contains crucial parameters for semiconducting, superconducting, and topological quantum electronic devices. The high mobility of holes and their low effective mass promote the confinement of spins in low-disorder Ge quantum dots by uniform potential landscapes\cite{Hendrickx2018gate}. Holes in Ge have large and tunable $g$-factors, with inherent strong spin-orbit interaction. These properties facilitate fast all-electrical qubit control\cite{watzinger_ge_2018}, qubit coupling at a distance via superconductors\cite{kloeffel2013circuit}, and are key ingredients for the emergence of Majorana zero modes for topological quantum computing.

Like Si, Ge can be isotopically purified into a nuclear spin-free material to achieve long spin lifetimes\cite{itoh_isotope_2014}.  In contrast, virtually every metal on Ge, including superconductors with high critical fields, show a Fermi level pinned close to the valence band\cite{dimoulas_fermi-level_2006}. This facilitates the injection of holes and thus the formation of Ohmic superconductor/semiconductor contacts, a key building block in hybrid quantum devices. 

These enticing prospects have motivated the theoretical framework for Ge-based spin-qubits\cite{terrazos2018light} and Majorana fermions\cite{maier2014majorana}. Experimental milestones in self-assembled Ge nanostructures include gate-tunable superconductivity
in Ge/Si nanowires\cite{xiang2006ge} and the demonstration of electrically driven spin qubits\cite{watzinger_ge_2018} and single-shot readout of single spins\cite{vukuvsic2018single} in Ge hut wires. 
\begin{figure}[H]
	\includegraphics[width=\linewidth]{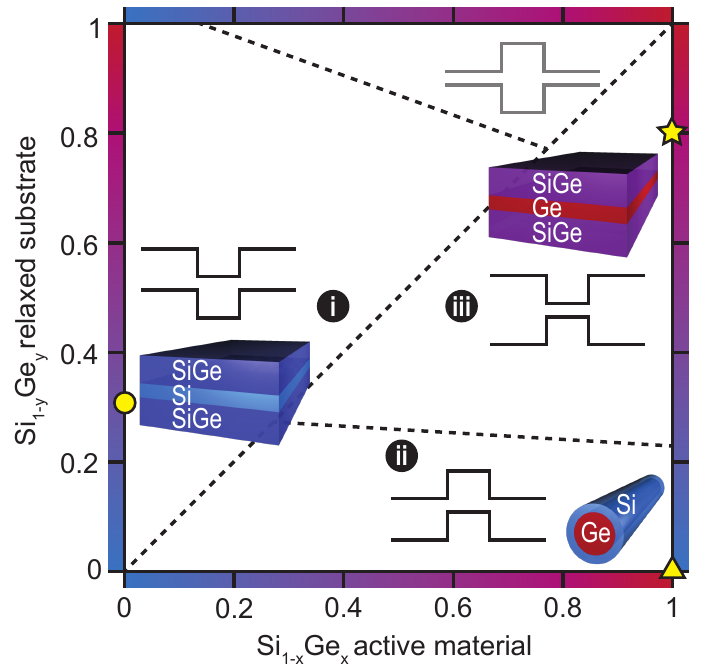}%
\caption{Schematics of SiGe heterostructures band-edge profiles as a function of the Ge concentration \textit{x} and \textit{y} in the active material and in the relaxed substrate, respectively. Star, circle, and triangle refer to Ge/SiGe, Si/SiGe and Ge/Si heterostructures, respectively. Adapted from Ref.\cite{virgilio2006type}}
\label{fig:play} 
\end{figure}

\begin{figure*}
	\includegraphics[width=\linewidth]{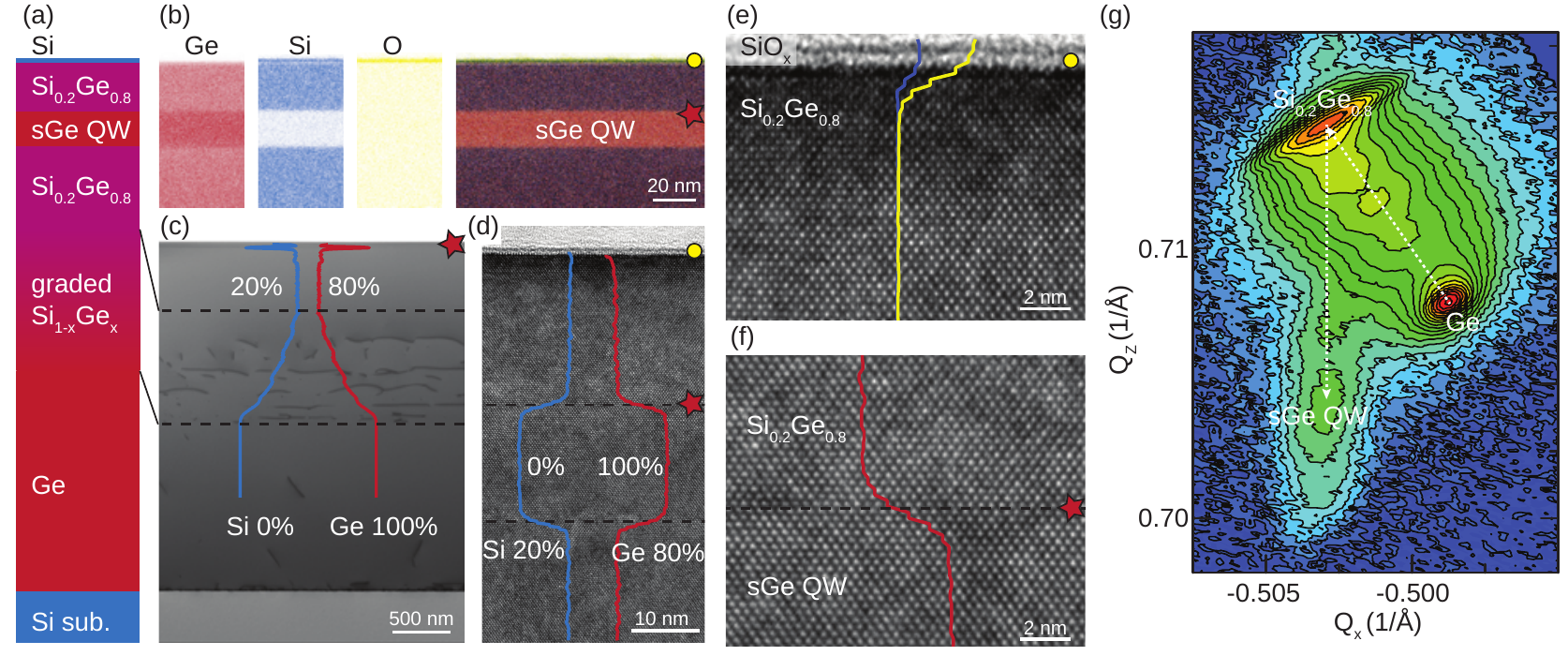}
	\caption{Structural characterisation of a Ge/SiGe heterostructure. (a) Layer schematics. (b) Ge, Si, and O signals from transmission electron microscopy with energy dispersive X-Ray analysis (STEM/EDX) of the Ge quantum well and nearby Si$_{0.2}$Ge$_{0.8}$; (c) STEM and (d-f) TEM showing the SiGe virtual substrate, quantum well, quantum well/barrier interface, barrier/surface interface. (g) X-ray diffraction reciprocal space map of (-2-24) reflection.}
\label{fig:str}
\end{figure*}
Recently, Ge/SiGe heterostructures have emerged as a planar technology that can bring together low disorder, potential for fast qubit driving, and avenue for scaling due to the compatibility with large scale manufacturing.
In Ge/SiGe, the band-edge profiles between compressively strained Ge and relaxed Ge-rich Si$_{1-y}$Ge$_y$  (Fig. \ref{fig:play}; star, $y\approx0.8$) produce a type I band alignment\cite{virgilio2006type}. This is different from Si/SiGe heterostructures (Fig. \ref{fig:play}; circle, $y\approx0.3$) and Ge/Si core-shell nanowires (Fig. \ref{fig:play}; triangle), where a type II band alignment promotes confinement of either electrons or holes, respectively. Charge carriers can populate the quantum well either by doping the heterostructure or via top gating. 
Holes confined in modulation doped Ge/SiGe have shown exceptionally high mobility of 1.5 million cm$^2$/Vs, strong spin orbit coupling\cite{failla_terahertz_2016}, and fractional quantum hall physics\cite{shi2015spinless}.

\begin{figure*}	\includegraphics[width=\linewidth]{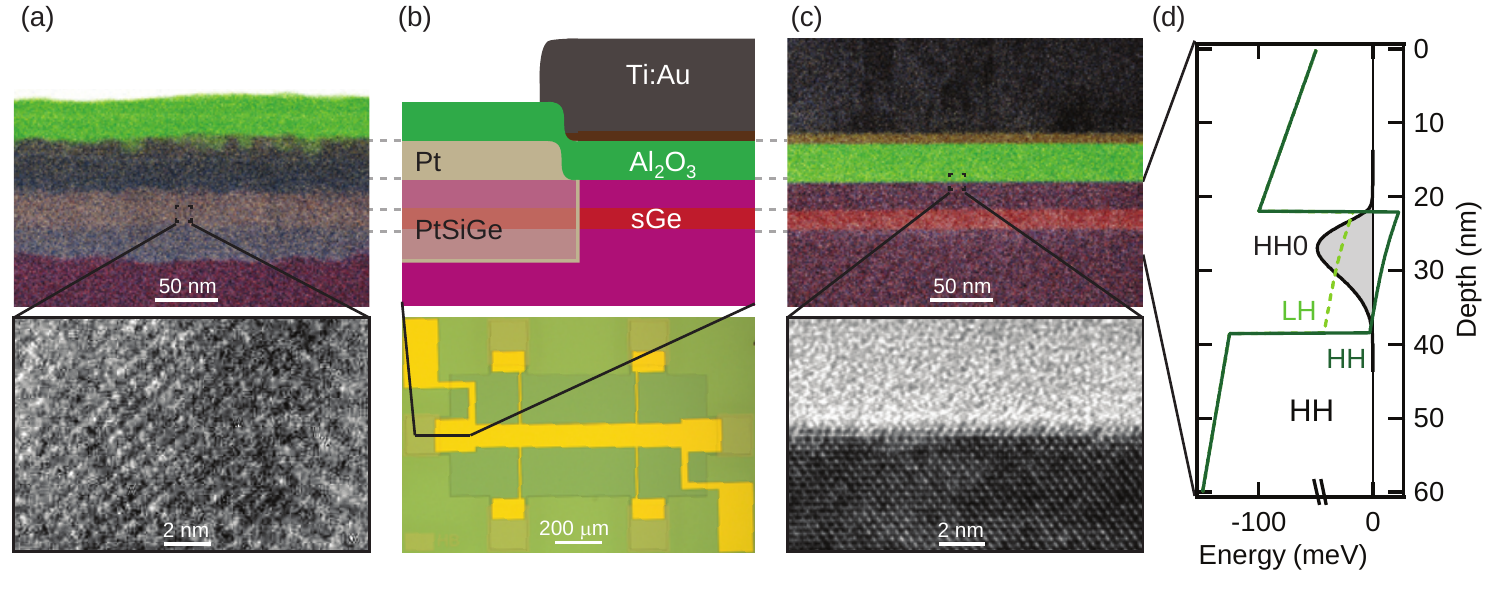}
\caption{Ge/SiGe heterostructure field effect transistor. (a) STEM/EDX (top) and TEM (bottom) in the ohmic contact region. (b) Device cross-section and optical image (upper and lower panels, respectively). (c) STEM/EDX (top) and TEM (bottom) in the channel region. (d) Bandstructure simulation with heavy holes (HH), light holes (LH) and charge distribution in the first subband HH0.}
\label{fig:dev}
\end{figure*}

In analogy to electron spin qubits in Si/SiGe\cite{maune2012coherent} it is preferable to completely eliminate dopant atoms, which are a major source of scattering, charge noise, and instability for the low-temperature operation of quantum devices\cite{borselli2011pauli}. Gate controlled quantum dots\cite{Hendrickx2018gate}, ballistic one dimensional channels\cite{mizokuchi2018ballistic}, and ballistic phase coherent superconductivity\cite{hendrickx2018ballistic} were demonstrated recently by using undoped Ge/SiGe. The added complexity in developing reliable gate-stacks compared to modulation-doped structures has limited the investigation of quantum transport properties in undoped Ge/SiGe so far\cite{laroche_magneto-transport_2016,su_effects_2017-1,hardy2018single}.

In this paper we demonstrate state of the art, very shallow, undoped Ge/SiGe heterostructures and devise a process for fabricating dopant-less heterostructure field effect transistors (H-FETs) without compromising on material quality. A comprehensive investigation of key electrical properties such as carriers mobility, critical density for conduction, effective mass, and g-factor, establishes Ge/SiGe as a promising platform for future hybrid quantum technologies.

\vspace{-0.25\baselineskip}
\section*{Results}
\noindent\textbf{Ge/SiGe heterostructures.} Figure \ref{fig:str} shows a schematic of the investigated undoped Ge/SiGe heterostructures along with the results of structural characterization to elucidate the crystallographic, morphological, and chemical properties of the stack. 

The Ge/SiGe heterostructure is grown on a 100 mm Si (001) substrate in a high-throughput reduced-pressure chemical vapour deposition reactor. The layer sequence comprises a Si$_{0.2}$Ge$_{0.8}$ virtual substrate obtained by reverse grading\cite{shah_reverse_2008,capellini2010strain}, a 16-nm-thick Ge quantum well, a 22-nm-thick  Si$_{0.2}$Ge$_{0.8}$ barrier, and a thin sacrificial Si cap (see Methods). The purpose of the Si cap is to provide a well-known starting surface for subsequent high-$\kappa$ metal gate stack deposition\cite{vincent2011si} and to possibly achieve a superior dielectric interface than what SiGe could offer. Both C and O concentration within the quantum well are below the secondary ion mass spectroscopy detection limit of ($4\times 10^{16}\text{ cm}^{-3}$) and ($8\times 10^{16}\text{ cm}^{-3}$), respectively, pointing to a very low impurity background level (see Supplementary Informations).

Figure \ref{fig:str}(c) highlights the crystalline quality of the Si$_{0.2}$Ge$_{0.8}$ virtual substrate. Defects and dislocations are confined to the lower layers, at the Si/Ge interface and in the graded Si$_{1-y}$Ge$_y$. As the Si (Ge) concentration in the SiGe alloy is increased (decreased), relaxation of the upper layers is promoted. By performing preferential etching (see Supplementary Informations) we estimate an upper bound for the threading dislocation density of $(3.0\pm0.5)\times10^7$ for the Si$_{0.2}$Ge$_{0.8}$. The Si and Ge concentrations profiles across the virtual substrate (Fig. \ref{fig:str}(c); blue and red curves respectively) confirm the achievement of linear reverse-graded SiGe with targeted alloy composition. 

In-plane and out-of plane lattice parameters are obtained from the X-Ray diffraction reciprocal space map (XRD-RSM)  in Fig. \ref{fig:str}(g). The Ge and Si$_{0.2}$Ge$_{0.8}$ buffer layers are over-relaxed compared to the Si substrate with a residual tensile strain of $\epsilon_{\parallel}=0.2\%$ and 0.26\%, respectively. This is typical in SiGe virtual substrates due to the difference in thermal contraction of the materials after cooling from the high growth temperature\cite{shah_reverse_2008,capellini2012high}. The peak corresponding to the Ge quantum well is vertically aligned to the peak of the Si$_{0.2}$Ge$_{0.8}$ buffer layer, indicating a pseudomorphic growth of the quantum well and resulting in an in-plane compressive strain of $\epsilon_{\parallel}=-0.63\%$. 

Figure \ref{fig:str}(b) shows the individual and combined signals of Si, Ge, and O signals from the strained Ge quantum well embedded between Si$_{0.2}$Ge$_{0.8}$. The Ge QW appears as a uniform layer of constant thickness and with sharp interfaces to the adjacent Si$_{0.2}$Ge$_{0.8}$. The increasing O and Si signals at the top of the heterostructure indicate that the Si cap has readily oxidized upon exposure to air. The absence of extended defects in the high resolution TEM images in Fig. \ref{fig:str}(d)-(f) indicates high crystalline quality in the quantum well and adjacent Si$_{0.2}$Ge$_{0.8}$. The high degree of control achieved in the deposition process results in the Si and Ge composition profiles in Fig. \ref{fig:str}(d)-(f), with constant Ge composition within each layer of the Si$_{0.2}$Ge$_{0.8}$/Ge/Si$_{0.2}$Ge$_{0.8}$ structure.

\begin{figure*}%
	\includegraphics[width=\linewidth]{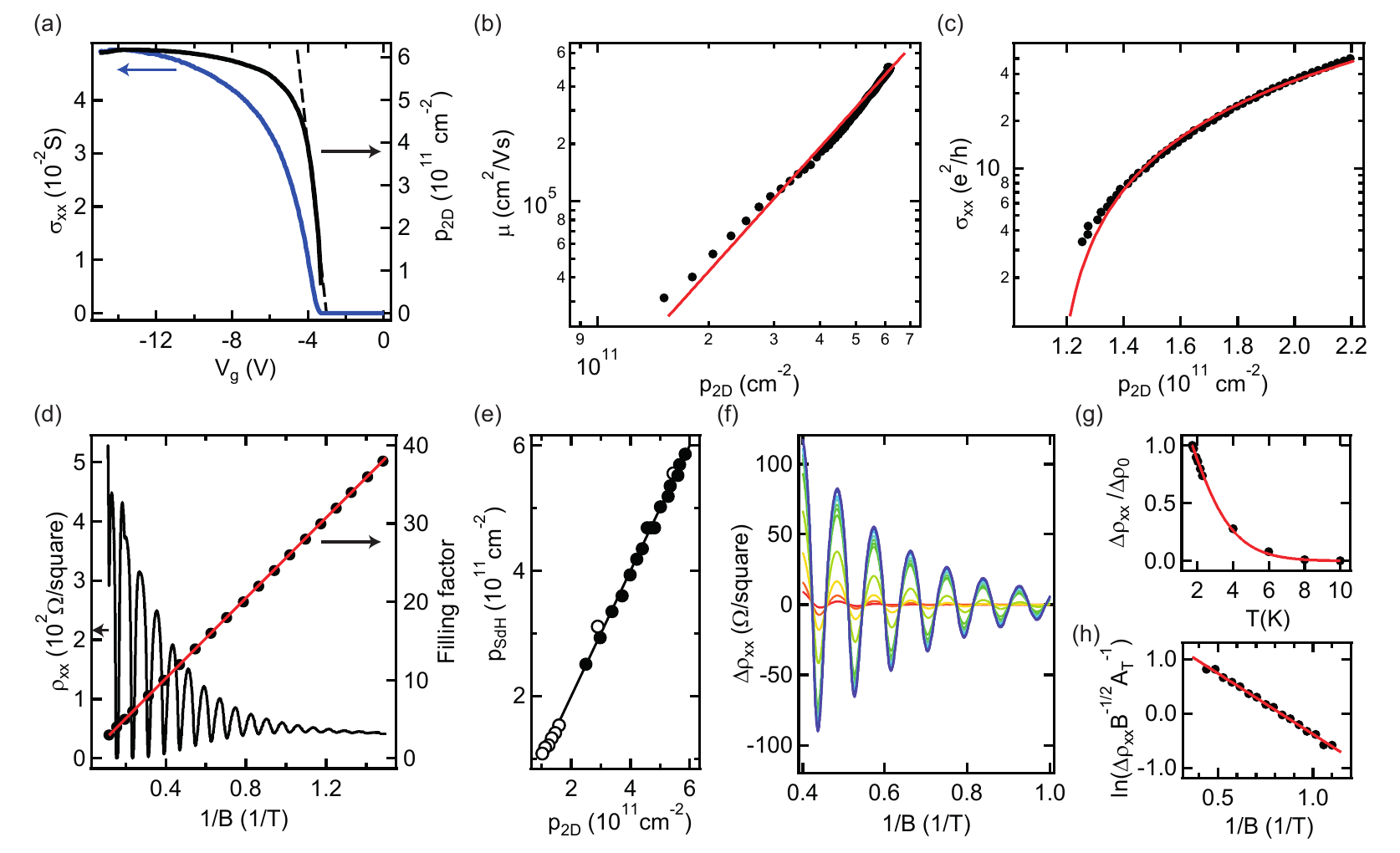}%
	\caption{Transport measurements at $T=1.7$ K as a function of magnetic field $B$ and carrier density $p_{2D}$. (a) Conductivity $\sigma_{xx}$ (blue line) and Hall density $p_{2D}$ (black line) as a function of gate voltage $V_g$ at 1.7 K. Dashed line is a linear fit of the gate-dependent density at low $V_g$. (b) Density dependent mobility $\mu$ (solid circles) and power law fit (red curve) (c) Density dependent $\sigma_{xx}$ (solid circles) and fit to percolation theory (red line). (d) Magnetoresistivity $\rho_{xx}$ (black line) and filling factor (solid circles) at saturation density as a function of inverse magnetic field $1/B$. Red line is the fit to the filling factor progression. (e) Density obtained by the analysis of the Shubnikov de Haas oscillations $p_{SdH}$ plotted against the corresponding Hall density $p_{2D}$. Open and solid circles are measured at $1.7$ K and $50$ mK, respectively. (f) Temperature dependence of the SdH oscillations $\Delta\rho$ in the range $T=$ 1.7-10 K, after background subtractions. (g) $\Delta\rho$ (solid circles) as a function of $T$, normalized at $\Delta\rho_0=\Delta\rho(T_0=1.7\text{K})$. The red line is the fit used to extract $m^{*}$. (h) Dingle plot at $T_0=1.7$ K (see Methods). The red line is the linear fit used to extract $\tau_q$.}
\label{fig:cla}
\end{figure*}

\noindent\textbf{Heterostructure field effect transistors.} Hall-bar shaped heterostructure field effect transistors (H-FETs) were fabricated to investigate the magnetotransport properties of the undoped Ge/SiGe. An external electric field is applied to the insulated top-gate thereby populating the Ge quantum well and creating a two-dimensional hole gas (2DHG). Compared to undoped Si/SiGe H-FETs\cite{mi2015magnetotransport}, we adopted a low-thermal budget, implantation-free process to obtain direct contact between diffused Pt metallic ohmics and the induced 2DHG (see Methods). This approach is possible due to the very low platinum germanosilicide hole Schottky barrier height\cite{kittl2008silicides,alptekin2009platinum}. 

The optical image of the final device is displayed in Fig. \ref{fig:dev}(b), together with the schematics of the transistor cross section at the gate/ohmic contact overlap region. The upper panels in Fig. \ref{fig:dev}(a) and (c) show STEM/EDX analysis in the ohmic contact region and under the top-gate, respectively. In the ohmic region (Fig. \ref{fig:dev}(a), top panel) Pt diffuses inside the SiGe barrier and surpasses the quantum well. A PtSiGe alloy is formed, with a Ge concentration less than the value of 0.8 in the as-grown material, due to the Pt dilution within. The formation of a PtSiGe alloy is also supported by the presence of crystalline grains, as visible by TEM (Fig. \ref{fig:dev}(a), bottom panel). Since Pt diffusion is achieved at significantly lower temperature than the quantum well growth  - 300  $^\circ$C vs 500  $^\circ$C, respectively - the crystalline quality of the heterostructure under the gate-stack is preserved. The critical Ge/SiGe interfaces after device processing (Fig. \ref{fig:dev}(c), top panel) are as sharp as in the as grown material (Fig. \ref{fig:str}(c)). Furthermore, the high-resolution TEM image in the bottom panel of Fig. \ref{fig:dev}(c) highlights the atomically flat semiconductor/oxide interface.
 
Figure \ref{fig:dev}(d) shows the band-structure in the H-FET at a carrier density of $1\times 11^{10}\text{ cm}^{-2}$ by solving the Schroedinger-Poisson equation as a function of the applied electric field at low temperatures. States in the quantum well with heavy hole (HH) symmetry are favored compared to light holes (LH) states, with the HH and LH band-edges split in energy by 40 meV. The wave-function of the fundamental HH state (HH0) is well confined in the Ge quantum well, with an energy splitting between the HH0 and LH0 states of 47 meV, and between HH0 and the first excited HH state (HH1) of 15 meV. These obtained energy splittings in Ge/SiGe are more than one order of magnitude larger than the valley splitting in the conduction band of Si/SiGe or Si/SiO$_2$ systems\cite{zwanenburg2013silicon}, supporting the possibility of obtaining well defined qubits in this material platform.

\noindent\textbf{Mobility, critical density, and effective mass.} Magnetotransport characterisation of the Ge/SiGe H-FETs was performed at low temperatures to elucidate the quantum transport properties of the 2DHG. The device is operated in accumulation mode, in which carriers populate the quantum well by applying a negative DC voltage bias ($V_g$) to the top gate. Upon applying a fixed AC voltage bias to source and drain contacts ($V_{sd}$), standard four-probe lock-in techniques allow to measure  the longitudinal and transverse components of the resistivity tensor ($\rho_{xx}$, $\rho_{xy}$, respectively), from which longitudinal ($\sigma_{xx}$) and transverse ($\sigma_{xy}$) conductivity are extracted. The active carrier density $p_{2D}$ is measured by the Hall effect and, consequently, the carrier mobility $\mu$ (see Methods). 

\begin{figure*}%
	\includegraphics[width=120mm]{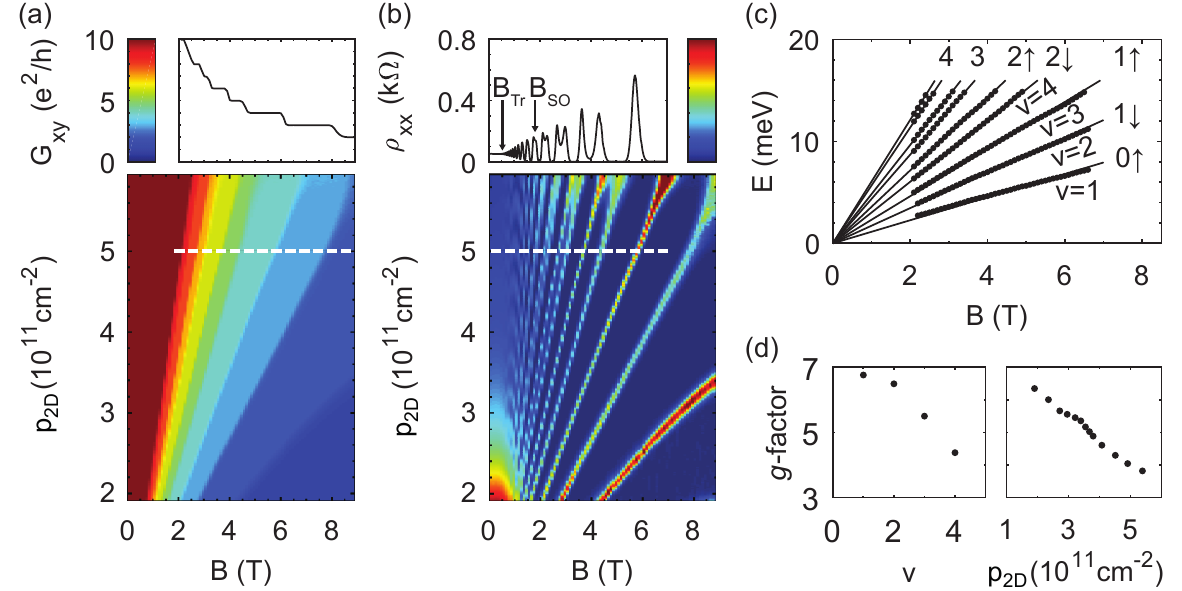}%
	\caption{Transport measurements at $T=50$ mK as a function of magnetic field $B$ and carrier density $p_{2D}$. (a) Quantized Hall conductance and (b) Shubnikov de Haas oscillations. (c) The local Fermi energy $E$ for different Landau levels. (d) Effective g-factor as a funcion of filling factor (left panel) and density (right panel).}
\label{fig:Fig6}
\end{figure*}
Figure \ref{fig:cla}(a) shows the conductivity and the carrier density as a function of gate voltage (blue and black curves, respectively). At zero applied $V_g$ there are no carriers induced in the quantum well, the undoped heterostructure is insulating, and no conduction is measured. Above a threshold bias ($V_g$ = -3.8 V), current starts flowing in the channel and $\sigma_{xx}$ increases monotonically until saturation. After turn-on, $p_{2D}$ increases linearly as $V_g$ sweeps more negative. This is consistent with a simple parallel-plate capacitor model in which the Si$_{0.2}$Ge$_{0.8}$ and Al$_2$O$_3$ layers are the dielectric layers between the Ge quantum well and the metallic top-gate. The extracted effective gate capacitance obtained by fitting the linear portion of the $p_{2D}$ $vs.$ $V_g$ curve, however, is reduced to $\approx60\%$ of the expected value. At larger $V_g$ the $p_{2D}$ $vs.$ $V_g$ curve deviates from linearity and eventually $p_{2D}$ saturates at a value of $6\times 10^{11}\text{ cm}^{-2}$. 

Figure \ref{fig:cla}(b) shows the density-dependent mobility. The mobility increases strongly with density over the entire range of accessible densities, without signs of saturation. By fitting the data to a power law dependence $\mu = p_{2D}$$^\alpha$ we find a large exponent $\alpha = 2.1$. Including local field corrections\cite{dolgopolov2003remote,gold2010mobility}, exponents ($\alpha \geq$ 1.5) indicate that the mobility is still limited by scattering from remote impurities at the dielectric/semiconductor interface, as seen previously in Si/SiGe heterostructures\cite{lu2011upper,laroche_magneto-transport_2016}. At saturation density $p_{2D}=6\times 10^{11}\text{ cm}^{-2}$  we measure a maximum mobility of $5\times 10^{5}\text{ cm}^2$/Vs, corresponding to a mean free path of $\approx6 \mu m$, setting new benchmarks for holes in shallow FET devices. 

The critical density for establishing metallic conduction in the channel is extracted by fitting the density-dependent conductivity (Fig. \ref{fig:cla}(c)) in the low density regime to percolation theory\cite{tracy2009observation,kim_lyon_2017}  $\sigma \propto A(p-p_p)^p$ cm$^{-2}$.. By fixing the exponent $p=1.31$, as expected in a 2D system, we estimate a percolation threshold density $p_p = 1.15\times10^{11}$ cm$^{-2}$ at 1.7 K, which sets an upper bound for $n_p$ at the temperature at which qubits typically operate ($T\leq$ 100 mK). Note that this value is in agreement with the qualitative estimate $p_c = 1.19\times10^{11}$ cm$^{-2}$ obtained by extrapolating to zero the linear region of mobility $\mu$ vs. log($p_{2D}$) curve above the critical density \cite{mccamey2005donor}. The obtained percolation threshold density is similar to the values reported in high quality Si MOSFET\cite{tracy2009observation,kim2017annealing}, and $\approx 2 \times$ higher than the values reported in undoped Si/SiGe\cite{mi2015magnetotransport}, possibly because the dielectric interface in our samples is much closer to channel (22 nm compared to 50 nm in Ref\cite{mi2015magnetotransport}).

Figure \ref{fig:cla}(d) shows the Shubnikov de Haas (SdH) oscillations in the magnetoresistivity at saturation density. The oscillations minima reach zero value at high fields and are periodic against the inverse of magnetic field $1/B$. From the linear filling factor progression (Fig. \ref{fig:cla}(d), red circles) we extract a density $p_{SdH}=6.1\times 10^{11}\text{ cm}^{-2}$, in  agreement with the Hall data. The agreement $p_{SdH}=p_{2D}$ is verified throughout the range of investigated density and temperature (Fig. \ref{fig:cla}(e)).
Figure \ref{fig:cla}(f) shows the temperature dependent magnetoresistance oscillation amplitude $\Delta\rho$ at a density $p_{2D}=5.4\times 10^{11}\text{ cm}^{-2}$ as a function of the inverse of the magnetic field $1/B$, after the subtraction of a polynomial background. The hole effective mass $m^*$ is extracted by fitting the damping of the SdH oscillation amplitude with increasing temperature at $B = 1.4~$T (Fig. \ref{fig:cla}(g), see Methods). The obtained value $m^* = (0.090 \pm 0.002)m_e$, where $m_e$ is the free electron mass, is comparable to previous reports in Ge/SiGe.\cite{laroche_magneto-transport_2016,hardy2018single}. The quantum lifetime $\tau_q$ at 1.7 K is extracted by fitting the SdH oscillation amplitudes envelope\cite{coleridge1989low}. From the Dingle plot in Fig. \ref{fig:cla}(h), we estimate $\tau_q = 0.74~$ps, corresponding to a large Dingle ratio $\tau_t/\tau_q = 27$, where $\tau_t$ is the transport lifetime. The obtained large Dingle ratio confirms that the mobility is limited by scattering from charges trapped at the dielectric/heterostructure interface.

\noindent\textbf{Landau fan diagram and effective g-factor.} In Fig. \ref{fig:Fig6}(a) and (b) we show color plots of both $\rho_{xx}$ as well as the transversal Hall signal $\sigma_{xy}$ at $T=50$ mK, as a function of out-of-plane magnetic field $B$ and carrier density $p_{2D}$, as obtained from the low-field Hall data. In this Landau fan diagram, both the quantum Hall effect (Fig. \ref{fig:Fig6}(a)) and SdH-oscillations (Fig. \ref{fig:Fig6}(b)) fan out linearly towards higher field and density. Observation of a  Landau fan diagram of such quality is a sign of the very low disorder in the 2D channel. As in the measurements at 1.7 K (Fig. \ref{fig:cla}(f)), we do not observe beatings in the SdH oscillations, which prevent us to directly measure the strength of the spin-orbit interaction. However, we estimate an upper bound for spin-splitting in the 2DHG of 1.5 meV from the peak width of 1.5 T observed in the Fourier transform of $\rho_{xx}$ against 1/B. 

We estimate  the out-of-plane g-factor by analyzing the difference in slope of the spin-up and spin-down level of the same Landau level\cite{huang2013direct}. We convert the carrier density to an energy (Fig. \ref{fig:Fig6}(c)) by assuming a free electron dispersion $E_F=\pi\hbar^2p_{2D}/m^*$, with $\hbar$ Planck's constant and $m^*=0.09m_e$ the effective mass in the system. The difference between the spin split levels is directly given by the Zeeman splitting energy $E_Z=g\mu_BB$. We plot this derived $g$-factor in the left panel of Fig. \ref{fig:Fig6}(d) and observe that it decays from $g=7$ to $g=4$ as a function of the filling factor. Alternatively we compare the field strength at which SdH oscillations appear, $B_{Tr}$, to the field at which the spin-splitting occurs $B_{SO}$, such that the $g$-factor can be estimated by $g=\frac{2m_e}{m^*}\frac{1}{1+B_{SO}/B_{Tr}}$\cite{mi2015magnetotransport}. The analysis is shown in Fig. \ref{fig:Fig6}(d) (right panel), with the range of the $g$-factor being in good agreement with the fan diagram analysis and other reports in Ge\cite{drichko2018effective}. The observed enhancement of the $g$-factor with decreasing densities has been previously observed and attributed to an increase in the strength of the effective Coulomb interaction\cite{sadofyev2002large,lu2017effective} or to the non-parabolicity of the valence band\cite{drichko2018effective}. 

\section*{Discussion}
The observed  $p_{2D}-V_g$ and $\mu-p_{2D}$ dependences are in line with previous studies on shallow undoped Si and Ge/SiGe heterostructures \cite{laroche2015scattering, su_effects_2017-1}. At small electric fields, carrier tunneling can occur from the shallow Ge quantum well to defect states in the band-gap of the dielectric/SiGe interface. Whilst tunneling reduces the gate capacitance, the passivated surface impurities by tunneled carriers lead also to an enhanced mobility\cite{huang2014screening}. At higher electric fields, the Fermi level aligns with the valence band edge at the dielectric/SiGe interface. Population of the surface quantum well prevents, by screening, further carrier accumulation in the buried channel, which reaches saturation. Nevertheless, only the buried quantum well contributes to transport, since the surface quantum well carrier concentration is likely below the mobility edge threshold, which is typically high for a channel at the Al$_2$O$_3$/Si interface. This interpretation is supported by the matching densities $p_{SdH}$ and $p_{SdH}$ (Fig. \ref{fig:cla}(d)), with no beating observed in the SdH oscillations: only one high mobility subband contributes to the measured transport.

In conclusion, by measuring key transport metrics at low temperatures we have shown that shallow and undoped Ge/SiGe heterostructures are a promising low-disorder platform for Ge quantum devices.  
The reported half-million mobility sets new benchmarks for Si and Ge shallow-channel H-FETs\cite{huang2014screening,laroche2015scattering,su_effects_2017-1}, while even higher mobilities may be obtained by by further improving the semiconductor/dielectric interface. Possible avenues in these directions include the removal of the native silicon oxide layer prior to high-\textit{$\kappa$} dielectric deposition and/or post-metallization thermal anneals. A better quality semiconductor/dielectric interface should also improve the critical density, which is a critical metric for quantum devices. 
The obtained effective mass of 0.09m$_{e}$ is much lighter than the 0.19m$_{e}$ value for electron in silicon  and close to the value of 0.067m$_{e}$ for electrons in GaAs. A light effective mass is beneficial for spin qubits since it corresponds to larger energy level spacing in quantum dots and relaxes lithographic fabrication requirements due to a larger extent of the wavefunction. Further improvements towards achieving the even lighter theoretical value of 0.05m$_{e}$\cite{terrazos2018light} will include lowering the growth temperature of the quantum well to reduce Ge diffusion from the quantum well in the barrier, thereby improving the abruptness of the quantum well/barrier interface and preventing possible spilling of the wave-function in the SiGe barrier.

\section*{Methods}
\small \textbf{Heterostructure growth.} The Ge/SiGe heterostructure is grown in a high-throughput reduced-pressure chemical vapour deposition (RP-CVD) reactor (ASMI Epsilon 2000) in one deposition cycle using germane (GeH$_4$) and dichlorosilane (SiH$_2$Cl$_2$) as precursor gases. Starting with a 100 mm Si (001) substrate, a 1.4-$\mu$m-thick layer of Ge is grown using a dual-step process. An initial low-temperature (400 $^\circ$C) growth of a Ge seed layer is followed by a higher temperature (625 $^\circ$C) overgrowth of a thick relaxed Ge buffer layer. Cycle anneals at 800 $^\circ$C are performed to promote full relaxation of the Ge. The subsequent 900-nm-thick reverse-graded Si$_{1-y}$Ge$_y$ layer\cite{shah_reverse_2008} is grown at 800$^\circ$ C with $y$ changing from 1 to 0.8. The SiGe virtual substrate is completed by a Si$_{0.2}$Ge$_{0.8}$ buffer layer of uniform composition, which is initially grown at 800 $^\circ$C. For the final 160 nm, the growth temperature is lowered to match the growth temperature of the subsequent layers (500 $^\circ$C). In this way growth interruption for temperature equilibration is avoided at the critical quantum well/virtual substrate interface. The heterostructure is completed with the deposition of a 16 nm-thick Ge quantum well, a 22 nm-thick  Si$_{0.2}$Ge$_{0.8}$ barrier, and a thin ($<$2 nm) sacrificial Si cap.
\newline
\\
\small \textbf{Structural analysis.} 
X-ray diffraction measurements were performed with a 9 kW SmartLab diffractometer from Rigaku equipped with a Ge(400x2) crystal collimator and a Ge(220x2) crystal analyzer using CuK1 radiation. The asymmetric (-2-24) reflection was used for reciprocal space mapping to determine in-plane and out-of-plane lattice parameters.
Transmission electron microscope investigation was carried out using a FEI Tecnai Osiris. For energy dispersive X-ray spectroscopy (EDX), the TEM was operated in scanning TEM (STEM) mode. The beam energy for all TEM measurements was 200 keV.
\newline
\\
\small \textbf{Device fabrication.}
The process for undoped Ge/SiGe H-FETs comprises the deposition of metallic ohmics, a high-$\kappa$ dielectric, and a metallic top-gate. Ohmic pads are deposited on top of a mesa structure by e-beam evaporation of 60 nm of Pt. An HF dip is performed prior Pt deposition to etch the native oxide at the surface and ensure that the Pt film is in direct contact with the underlying Si$_{0.2}$Ge$_{0.8}$. The subsequent atomic layer deposition of 30 nm of Al$_2$O$_3$ at 300 $^\circ$C has the twofold purpose of electrically isolating the transistors top-gate from the channel as well as providing the thermal budget needed to drive the Pt ohmics in the Si$_{0.2}$Ge$_{0.8}$. Finally, the top-gate is realized by depositing 10/150 nm of Ti/Au.
\newline
\\
\small \textbf{Electrical measurements.} Magnetotransport data in the temperature range of 1.7 to 10K has been obtained in a $^4$He variable temperature insert refrigerator equipped with a $9$ T magnet. Magnetotransport data at 50 mK were obtained in a $^3$He dilution refrigerator equipped with a $9$ T magnet. A bias in the 0.1-1 mV range, frequency 7.7 Hz, is applied to the source and drain contacts.  The source drain current $I_{SD}$, the longitudinal voltage $V_{xx}$, and the transverse Hall voltage $V_{xy}$ are measured; the longitudinal resistivity $\rho_{xx}$ and transverse Hall resistivity $\rho_{xx}$ are calculated as $\rho_{xx}=V_{xx}/I_{SD}\times W/L$ and $\rho_{xy}=V_{xy}/I_{SD}$, respectively (aspect ratio $L/W=5$). Longitudinal ($\sigma_{xx}$) and transverse ($\sigma_{xy}$) conductivity are calculated from $\rho_{xx}$ and $\rho_{xy}$ by tension inversion. The electrically active Hall carrier density $p_{2D}$ is obtained from the linear dependence of the Hall resistivity with perpendicular magnetic field ($\rho_{xy}=B/p_{2D}e$) at low magnetic field values ($B \leq 0.5$ Tesla). The carrier mobility $\mu$ is obtained from the relationship $1/\rho_{xx}=p_{2D}e\mu$.
The effective mass is fitted from the damping of the SdH oscillations by using the expression\cite{de1993effective}:
\begin{equation}
\dfrac{\Delta\rho_{xx}}{\Delta\rho_{0}} = \dfrac{\Delta\rho / \rho_{0}~(T)}{\Delta\rho / \rho_{0}~(T_{0})} = \dfrac{A_{T}}{A_{T_{0}}} = \dfrac{T~sinh(\alpha T_{0})}{T_{0}~sinh(\alpha T)}~,
\end{equation}
where $\alpha = \dfrac{2\pi k_{B} m^*}{\hbar eB}$, $k_{B}$ is Boltzmann constant, $\hbar$ the Planck constant, $\rho_{0}$ is the zero field resistivity, and $T_{0} = 1.7K$ is the coldest temperature at which the oscillations were measured.
\newline

\section*{Additional information}
{\small \textbf{Competing interests:} The authors declare no competing interests.}

\section*{Author contributions}
{\small A.S.~fabricated the heterostructures and D.S.~fabricated the field effect transistors. D.S., L.Y. and M.L.~performed the magnetotransport characterisation, supported by G.S.. M.A.S. performed EDX-(S)TEM analysis, P.Z. did X-ray characterisation and B.P.W. and M.Vi. performed band-structure calculations. A.S, D.S, L.Y, N.W.H., G.C. and G.S.analysed the data. A.S, D.S., N.W.H. and G.S. ~wrote the manuscript with input from all authors. G.S.~conceived and supervised the project.}

\end{document}